\documentclass{article}
\usepackage{spconf,amsmath,graphicx,hyperref}
\usepackage{amssymb}
\usepackage{booktabs}
\usepackage{multirow}
\usepackage{multicol}
\usepackage{array,tabularx}
\newcolumntype{Y}{>{\centering\arraybackslash}X}
\newcolumntype{L}[1]{>{\raggedright\arraybackslash}p{#1}}
\usepackage{enumitem}

\title{Post-Training Zero-Shot TTS for Fine-Grained Emotion and Duration Control via Natural Language}
\name{
Lianru Gao$^{1,\dagger}$,
Yujie Guo$^{1,\dagger}$,
Yong Qin$^{1,*}$
}
\address{
$^1$TMCC, College of Computer Science, Nankai University, Tianjin, China\\
Email: 2211995@mail.nankai.edu.cn, qinyong@nankai.edu.cn
}
\makeatletter
\let\oldthebibliography\thebibliography
\renewcommand{\thebibliography}[1]{%
  \oldthebibliography{#1}%
  \setlength{\itemsep}{0pt}%
  \setlength{\parsep}{0pt}%
  \fontsize{9pt}{9.8pt}\selectfont%
}
\makeatother
\begin{document}
\ninept
\setlength{\parskip}{0pt}
\maketitle

\begingroup
\renewcommand{\thefootnote}{}
\footnotetext{$^{\dagger}$ Equal contribution.}
\footnotetext{ $^{*}$ Corresponding author.}
\endgroup

\begin{abstract}
Audiobook narration, conversational agents, and audiovisual dubbing require fine-grained control over emotion and pacing within an utterance, yet many text-to-speech (TTS) systems remain limited to utterance-level conditioning. We propose a unified post-training framework for natural-language segment-level emotion and duration control. Supervised fine-tuning (SFT) establishes instruction following, while group relative policy optimization (GRPO) further improves control accuracy using emotion, duration, content, and speaker rewards. Without modifying the pretrained architecture or adding inference-time control modules, our method improves fine-grained controllability while preserving intelligibility and speaker identity, demonstrating post-training as an effective way to extend pretrained TTS models.

\end{abstract}
\section{Introduction}
\label{sec:intro}

Practical speech synthesis requires not only intelligibility and speaker consistency, but also flexible control over expression and timing. Applications such as audiobook narration, conversational agents, and audiovisual dubbing often require emotion and pacing to vary across different segments within a single utterance. However, although recent large-scale TTS models offer increasingly rich control over emotion, style, and duration, such control is typically applied at the utterance level, making fine-grained segment-specific control difficult.

Natural-language instructions offer an intuitive interface for controllable TTS, but existing methods such as PromptTTS and InstructTTS mainly operate at the utterance level~\cite{guo2023prompttts,yang2024instructtts}. Fine-grained control has been explored through segment-, token-, or frame-level mechanisms for emotion, duration, and speaking rate~\cite{wang2026word,liang2026ted,mai2026magic,liu2026emotra}, but these approaches typically rely on structured control signals or specialized designs. Consequently, natural-language joint control of segment-level emotion and duration remains largely underexplored.

Meanwhile, post-training has emerged as an effective way to refine pretrained TTS models. Recent studies use Group Relative Policy Optimization (GRPO) or preference optimization to improve linguistic fidelity, emotion control, and complex instruction following~\cite{liu2026group,li2026emorl,zhou2026sequential,chen2026flexivoice}. These results show that post-training can effectively shape speech-generation behavior beyond supervised learning. However, joint segment-level emotion and duration control through natural-language instructions remains largely unexplored.

\begin{figure*}[t]
    \centering
    \includegraphics[width=0.95\textwidth,height=7cm]{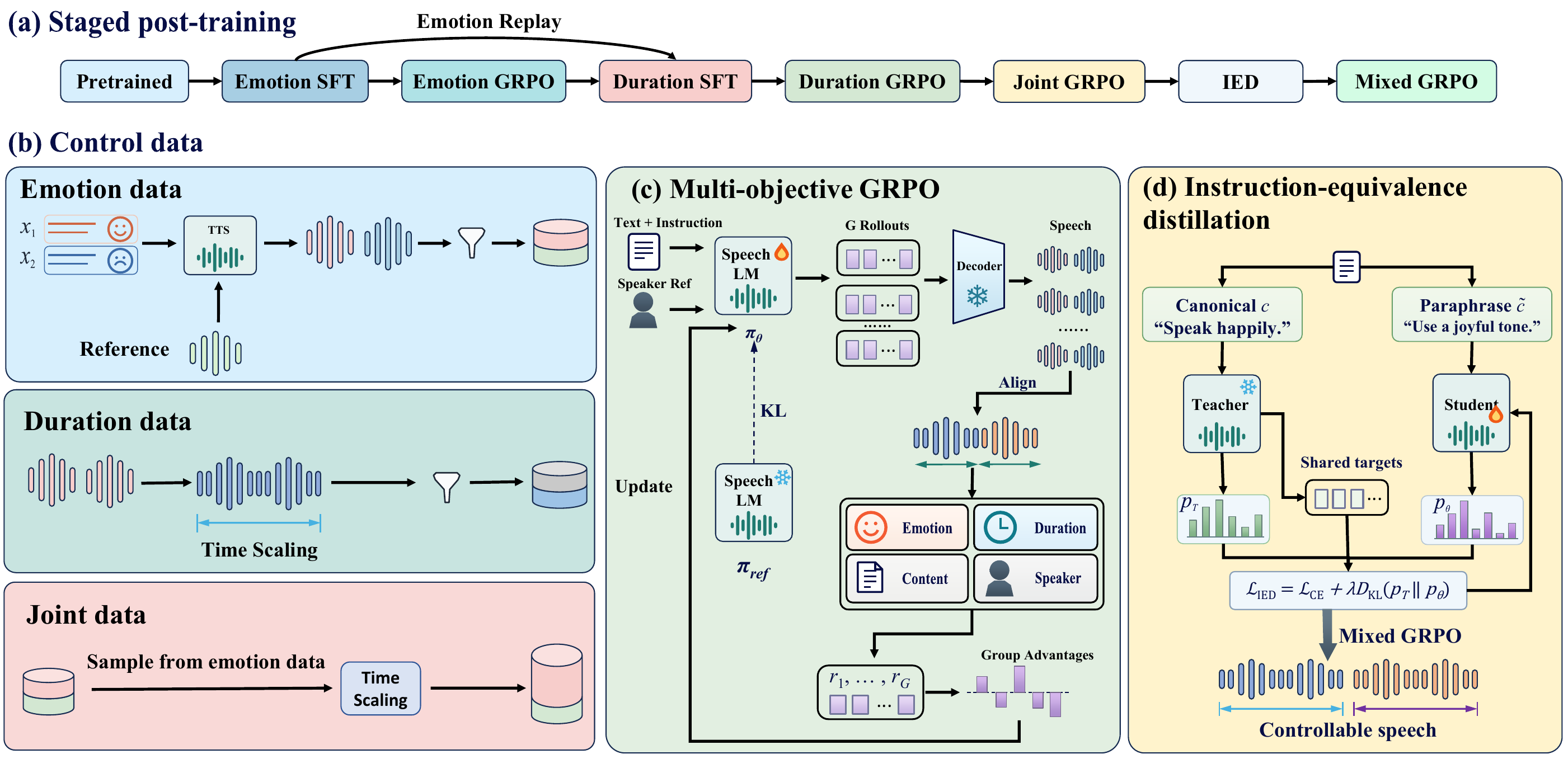}
    \caption{Overview of the proposed unified post-training framework for
    natural-language segment-level emotion and duration control.}
    \label{fig:framework}
    \vspace{-3mm}
\end{figure*}

Motivated by these observations, we investigate whether fine-grained controllability can be acquired through post-training rather than architectural modification. We first use supervised fine-tuning (SFT) to establish the correspondence between natural-language instructions and segment-level speech behaviors. However, fine-grained speech control is outcome-oriented: multiple acoustic realizations may satisfy the same instruction, while token-level supervision does not directly measure whether generated speech meets the requested emotion and duration of each segment. We therefore apply GRPO with segment-level emotion and duration rewards, alongside content and speaker-preservation objectives to maintain intelligibility and speaker identity. SFT establishes the instruction-following capability, while GRPO further refines fine-grained control without modifying the pretrained generation architecture.

A further challenge is the variability of natural-language instructions: semantically equivalent phrasings may yield inconsistent control when supervised training covers limited templates. Previous distillation methods transfer teacher-generated sequences~\cite{kim2016sequence} or improve TTS reliability through ASR verification~\cite{asaria2026reliable}. Inspired by these methods and to address instruction sensitivity, we introduce instruction-equivalence distillation, which aligns the output token distributions of a frozen teacher conditioned on a canonical instruction and a student conditioned on its paraphrase under a shared speech target. This encourages consistent generation across equivalent instructions and improves generalization to unseen phrasings. Our contributions are threefold:

\begin{itemize}
    \item We develop a unified post-training framework that achieves natural-language segment-level emotion and duration control using only a small amount of SFT data, with GRPO further refining control accuracy without additional inference-time modules.
    \item We introduce instruction-equivalence distillation to improve robustness to diverse natural-language expressions of the same control intent.
    \item We evaluate emotion-only, duration-only, and joint control using segment-level and strict utterance-level metrics, complemented by counterfactual and instruction-variation tests that examine the limits of instruction following.
\end{itemize}

\section{Method}
\label{sec:method}

Given $K$ text spans $x=(x_1,\ldots,x_K)$ and a natural-language instruction $c$ specifying each span's emotion, duration, or relative speaking rate, we aim to generate a single utterance satisfying these controls while preserving content and reference speaker identity. We fine-tune the speech language model of CosyVoice2~\cite{du2024cosyvoice}, freezing its speech tokenizer, acoustic generator, and vocoder. Figure~\ref{fig:framework} illustrates the training pipeline.

\subsection{Training Data Construction}

\noindent\textbf{Emotion data.}
We synthesize segments in five emotions using IndexTTS2~\cite{zhou2026indextts2}, with speaker references from Seed-TTS~\cite{anastassiou2406seed} and three manually screened ESD emotion references per category\cite{zhou2021seen}, selected using MERaLiON-SER-v1 confidence~\cite{sailor2025meralion}. The segments are synthesized separately, concatenated, and manually screened, yielding a 56.6-hour emotion training set.

\begin{table*}[t]
\caption{Objective evaluation of standard control.
ASR: WER/CER; Spk: speaker similarity;
JR: boundary jump rate; MER/E2V: emotion classifiers;
Acc/Utt: segment/strict utterance accuracy;
A15/U15: segment/strict duration accuracy;
RA/DA: range/direction accuracy.}
\label{tab:standard_control}
\renewcommand{\arraystretch}{1.01}
\centering
\resizebox{0.95\textwidth}{!}
{%
\begin{tabular}{l|ccc|cccc|cccc|cc}
\hline
\multicolumn{1}{c|}{\multirow{3}{*}{\textbf{System}}} & \multicolumn{3}{c|}{\multirow{2}{*}{\textbf{Basic Quality}}} & \multicolumn{4}{c|}{\textbf{Emotion}} & \multicolumn{4}{c|}{\multirow{2}{*}{\textbf{Duration}}} & \multicolumn{2}{c}{\multirow{2}{*}{\textbf{Joint}}} \\
\multicolumn{1}{c|}{} & \multicolumn{3}{c|}{} & \multicolumn{2}{c}{\textbf{MER↑}} & \multicolumn{2}{c|}{\textbf{E2V↑}} & \multicolumn{4}{c|}{} & \multicolumn{2}{c}{} \\
\multicolumn{1}{c|}{} & \textbf{ASR↓} & \textbf{Spk↑} & \textbf{JR↓} & \textbf{Acc} & \textbf{Utt} & \textbf{Acc} & \textbf{Utt} & \textbf{A15↑} & \textbf{U15↑} & \textbf{RA↑} & \textbf{DA↑} & \textbf{MER↑} & \textbf{E2V↑} \\ \hline
CosyVoice2 & 77.6 & 0.792 & \textbf{0.119} & 26.7 & 3.5 & 25.3 & 1.2 & 25.8 & 6.2 & 15.0 & 37.1 & 0.0 & 0.0 \\
CosyVoice3 & 11.1 & \textbf{0.806} & 0.124 & 31.6 & 5.3 & 32.7 & 7.7 & 44.9 & 18.6 & 23.2 & 41.2 & 1.8 & 0.6 \\
IndexTTS2 & \textbf{3.4} & 0.768 & 0.130 & 28.4 & 3.5 & 29.8 & 4.7 & - & - & - & - & - & - \\
TED-TTS & 7.0 & 0.627 & 0.138 & 51.8 & 19.4 & 46.9 & 14.7 & 46.9 & 16.5 & 26.3 & 49.0 & 1.2 & 0.6 \\
MAGIC-TTS & 29.2 & 0.758 & 0.130 & - & - & - & - & 46.4 & 21.7 & 22.2 & 46.9 & - & - \\ 
Joint-RL & 4.1 & 0.795 & 0.131 & \textbf{77.1} & \textbf{54.1} & \textbf{56.4} & \textbf{24.1} & \textbf{54.6} & \textbf{32.0} & \textbf{35.6} & \textbf{59.8} & \textbf{5.3} & \textbf{1.8} \\ \hline
\end{tabular}
}
\vspace{-4.5mm}
\end{table*}

\noindent\textbf{Duration and joint-control data.} We construct duration supervision from transcribed LibriTTS-R~\cite{koizumi2023libritts} recordings by independently stretching or compressing same-speaker segments using pitch-preserving PSOLA and concatenating them. Measured durations and applied ratios define exact-duration and relative-rate instructions, yielding distinct timing targets for the same text. Although duration-only SFT and GRPO use English data exclusively, the control generalizes to Chinese evaluation sets. We apply this procedure to synthesized emotional speech, preserving emotion labels while adding timing constraints for joint supervision. The duration-only and joint-control sets contain 168 and 30 hours, respectively.

\subsection{Emotion-Control Post-Training}

\noindent\textbf{Supervised initialization.} Given target speech tokens $z^{*}=(z_1^{*},\ldots,z_T^{*})$, we train the speech-token policy with teacher-forced cross-entropy $\mathcal{L}_{\mathrm{CE}}$ conditioned on $(x,c)$. The instruction serves only as conditioning context and is not part of the spoken target.

\noindent\textbf{Emotion-focused GRPO.}
Starting from the SFT model, we further optimize emotion control with GRPO~\cite{shao2024deepseekmath}. For each instruction, we sample multiple speech candidates, locate controlled segments with Qwen3-ForcedAligner~\cite{shi2026qwen3}, and predict their emotions using MERaLiON-SER-v1~\cite{sailor2025meralion}. The emotion-control reward is defined as
\begin{equation}
R_{\mathrm{ctrl}}
=
\lambda_{\mathrm{conf}}R_{\mathrm{conf}}
+
\lambda_{\mathrm{seg}}R_{\mathrm{seg}}
+
\lambda_{\mathrm{utt}}R_{\mathrm{utt}},
\end{equation}
where $R_{\mathrm{conf}}$, $R_{\mathrm{seg}}$, and $R_{\mathrm{utt}}$ denote emotion confidence, segment accuracy, and strict utterance accuracy, respectively.

To preserve linguistic content and speaker identity, we further incorporate ASR and speaker rewards:
\begin{equation}
R
=
R_{\mathrm{ctrl}}
+
\lambda_{\mathrm{asr}}R_{\mathrm{asr}}
+
\lambda_{\mathrm{spk}}R_{\mathrm{spk}}.
\end{equation}
Here, $R_{\mathrm{asr}}$ and $R_{\mathrm{spk}}$ encourage content and speaker preservation. GRPO optimizes group-relative advantages with KL regularization to the reference policy.

\subsection{Duration Acquisition and Joint Refinement}

\noindent\textbf{Duration SFT.} We fine-tune the emotion-refined model on duration-control data, replaying emotion examples to mitigate forgetting. The objective combines speech-token prediction with auxiliary supervision of segment token lengths and relative boundaries:
\begin{equation}
\mathcal{L}_{\mathrm{dur\text{-}SFT}} = \mathcal{L}_{\mathrm{CE}} + \lambda_{\mathrm{len}}\mathcal{L}_{\mathrm{len}} + \lambda_{\mathrm{bnd}}\mathcal{L}_{\mathrm{bnd}}.
\end{equation}
Auxiliary smooth-$L_1$ losses supervise log token counts and cumulative normalized boundaries via a head discarded at inference.

\noindent\textbf{Duration-focused GRPO.} Qwen3-ForcedAligner~\cite{shi2026qwen3} aligns generated speech with text to measure segment durations. Exact-duration requests use a smooth score
\begin{equation}
s_k=\exp(-|\log(\hat d_k/d_k)|/\tau_d),
\end{equation}
where $d_k$ and $\hat d_k$ are target and generated durations. Relative-rate requests are scored by interval compliance and proximity to target ratios. The duration reward combines soft scores, segment accuracy, strict utterance accuracy, and cumulative boundary agreement:
\begin{equation}
R_{\mathrm{dur}} = \lambda_{\mathrm{soft}}R_{\mathrm{soft}} + \lambda_{\mathrm{seg}}R_{\mathrm{seg}} + \lambda_{\mathrm{utt}}R_{\mathrm{utt}} + \lambda_{\mathrm{bnd}}R_{\mathrm{bnd}}.
\end{equation}

Exact-duration and relative-rate requests use relative-error and interval criteria, respectively, with strict accuracy requiring all controlled segments to pass. Timing variants share text and speaker but use separate GRPO groups per instruction, promoting instruction-dependent pacing without cross-instruction reward normalization. ASR and speaker rewards remain active.

\textbf{Joint refinement.}
We mix emotion-only, duration-only, and joint-control examples. Content and speaker rewards apply universally, while emotion and duration rewards are activated only when the corresponding control is present, with task-specific weights. This stage restores emotion control while preserving timing control.

\subsection{Instruction-Equivalence Distillation}

Starting from the joint-GRPO checkpoint, a frozen teacher generates speech-token targets $z^{*}$ under the canonical instruction $c$, while the student uses an intent-preserving paraphrase $\tilde c$, sharing the same text and teacher-generated token history. The student is optimized via

\begin{equation}
\mathcal{L}_{\mathrm{IED}}
=
\mathcal{L}_{\mathrm{CE}}(z^{*}\mid x,\tilde c)
+
\frac{\lambda_{\mathrm{KL}}}{T}
\sum_{t=1}^{T}
D_{\mathrm{KL}}
\left(
 p_T^{\,t}
\|
 p_{\theta}^{\,t}
\right),
\end{equation}
where $p_T^{\,t}$, $p_{\theta}^{\,t}$ denote the teacher and student token distributions, respectively. This objective encourages consistent generation across equivalent instructions.

Finally, we apply a short GRPO stage to the distilled model using mixed emotion-only, duration-only, and joint-control examples to restore and further improve control accuracy.

\section{Experiments}
\label{sec:experiments}

\subsection{Evaluation Setup}

Our test set comprises 534 examples disjoint from post-training data: 170 emotion-only, 194 duration-only, and 170 joint-control. Emotion texts are drawn from held-out MED-TTS, while duration samples use LibriTTS-R and Chinese MED-TTS at an English-to-Chinese ratio of approximately 1:2, with exact-duration or qualitative-rate instructions.

The emotion counterfactual set contains 50 Chinese/English-balanced neutral-text groups, each with five emotion requests and one no-control probe (250 requests, 50 probes). The duration counterpart contains 50 groups, each with three rate patterns and one probe (150 requests, 50 probes), using targets from held-out recordings and time-scaled variants. Text, speaker reference, and seed are fixed within groups. Both sets are disjoint from training at text and segment levels, with unseen evaluation speakers; probes are excluded from controlled-request metrics. The instruction-variation set paraphrases the 534 requests while fixing texts, references, and targets to assess wording robustness.

We compare CosyVoice2~\cite{du2024cosyvoice}, CosyVoice3~\cite{du2025cosyvoice}, and IndexTTS2~\cite{zhou2026indextts2} via native natural-language interfaces, and TED-TTS~\cite{liang2026ted} and MAGIC-TTS~\cite{mai2026magic} via structured controls. IndexTTS2 lacks natural-language segment-level duration control and is excluded from duration evaluation; MAGIC-TTS supports only duration/pause control and is excluded from emotion evaluation. TED-TTS uses segment-wise emotion descriptions and duration tokens. Structured-control instruction-variation scores are treated as references, not wording-robustness measures. All systems share identical texts, speaker references, targets, and scoring.

\subsection{Objective Metrics}
\label{sec:objective_metrics}

Qwen3-ForcedAligner~\cite{shi2026qwen3} locates controlled segments, whose emotions are evaluated by MERaLiON-SER-v1 (MER) and emotion2vec+ large (E2V)~\cite{ma2024emotion2vec}. E2V serves as an independent evaluator since MER provides the training reward. We report segment accuracy (Acc), strict utterance accuracy (Utt, all segments correct), and dual-classifier accuracy.

For duration control, Acc@15 (A15) measures the fraction satisfying $|\hat d_k-d_k|/d_k\leq0.15$, while strict utterance Acc@15 (U15) requires all segments to pass. For qualitative control, range accuracy (RA) checks request-specific intervals, while direction accuracy (DA) requires $r_k=\hat d_k/d_k^{\mathrm{ref}}$ to be $<0.95$, within $[0.95,1.05]$, or $>1.05$ for fast, normal, and slow requests. Joint accuracy requires correct emotion and duration control with ASR error $\leq20\%$. SenseVoice~\cite{an2024funaudiollm} WER/CER and CAM++~\cite{wang2023cam++} cosine similarity assess content and speaker preservation.

Following EmoTra-TTS~\cite{liu2026emotra}, we detect F0 jumps on 10-ms frames using $a_t=|f_{t+1}-2f_t+f_{t-1}|$ and $\tau=\operatorname{median}(a)+3\times1.4826\operatorname{MAD}(a)$. Boundary jump rate (JR-F0) measures the proportion of jumps within $\pm0.5$ s of aligned segment boundaries.



\subsection{Human Evaluation}
We conduct two human listening tests to complement objective metrics, with three raters per sample. First, raters label emotions for segments where the classifiers disagree, examining classifier-specific errors and potential overfitting to the reward classifier. Second, raters assess boundary and overall naturalness by comparing our single-pass joint-control model with CosyVoice3 and IndexTTS2 baselines that independently synthesize and concatenate segments, evaluating transition smoothness and utterance naturalness.

\section{Results}


\subsection{Control Performance.}
Table~\ref{tab:standard_control} compares Joint-RL with existing TTS systems. Joint-RL achieves superior emotion and duration control, particularly under strict utterance-level evaluation, outperforming structured-control systems using only natural-language instructions while preserving content fidelity and speaker similarity. CosyVoice2 tends to verbalize control instructions, degrading content accuracy. Nevertheless, satisfying emotion, duration, and content constraints simultaneously remains challenging.

\begin{table*}[t]
\caption{Ablation study of post-training stages}
\label{tab:standard_control2}
\renewcommand{\arraystretch}{1.01}
\centering
\resizebox{0.95\textwidth}{!}
{%
\begin{tabular}{l|ccc|cccc|cccc|cc}
\hline
\multicolumn{1}{c|}{\multirow{3}{*}{\textbf{System}}} & \multicolumn{3}{c|}{\multirow{2}{*}{\textbf{Basic Quality}}} & \multicolumn{4}{c|}{\textbf{Emotion}} & \multicolumn{4}{c|}{\multirow{2}{*}{\textbf{Duration}}} & \multicolumn{2}{c}{\multirow{2}{*}{\textbf{Joint}}} \\
\multicolumn{1}{c|}{} & \multicolumn{3}{c|}{} & \multicolumn{2}{c}{\textbf{MER↑}} & \multicolumn{2}{c|}{\textbf{E2V↑}} & \multicolumn{4}{c|}{} & \multicolumn{2}{c}{} \\
\multicolumn{1}{c|}{} & \textbf{ASR↓} & \textbf{Spk↑} & \textbf{JR↓} & \textbf{Acc} & \textbf{Utt} & \textbf{Acc} & \textbf{Utt} & \textbf{A15↑} & \textbf{U15↑} & \textbf{RA↑} & \textbf{DA↑} & \textbf{MER↑} & \textbf{E2V↑} \\ \hline
Emotion-SFT & 35.2 & 0.794 & 0.124 & 55.3 & 22.4 & 42.7 & 8.8 & 24.7 & 6.2 & 13.4 & 37.1 & 0.0 & 0.0 \\ 
Emotion-RL & 39.0 & 0.780 & 0.130 & \textbf{89.8} & \textbf{77.1} & \textbf{68.9} & \textbf{39.4} & 24.2 & 4.1 & 10.3 & 36.1 & 1.8 & 1.2 \\
Duration-SFT & 3.9 & 0.805 & 0.136 & 56.2 & 23.5 & 41.6 & 8.8 & 34.0 & 12.4 & 30.9 & 50.0 & 2.9 & 1.2 \\
Duration-RL & 3.8 & \textbf{0.806} & \textbf{0.123} & 53.3 & 20.6 & 40.4 & 8.2 & 52.1 & 26.8 & 28.9 & 62.4 & 1.2 & 0.6 \\
Joint-RL & 4.1 & 0.795 & 0.131 & 77.1 & 54.1 & 56.4 & 24.1 & \textbf{54.6} & \textbf{32.0} & \textbf{35.6} & 59.8 & \textbf{5.3} & \textbf{1.8} \\
IED+GRPO & \textbf{3.7} & 0.793 & 0.128 & 79.6 & 58.2 & 56.7 & 22.9 & 46.4 & 23.7 & 31.4 & \textbf{63.4} & \textbf{5.3} & 1.2 \\ \hline
\end{tabular}
}
\vspace{-5.5mm}
\end{table*}

\subsection{Ablation Study.}
Table~\ref{tab:standard_control2} examines the effects of successive post-training stages on control performance and speech quality. Emotion-SFT establishes basic emotion control, while Emotion-RL substantially improves accuracy, highlighting the importance of reinforcement learning for fine-grained control. Duration-SFT restores content fidelity and introduces timing control, with Duration-RL further improving precision. However, emotion accuracy declines substantially, revealing interference between control objectives. Joint-RL mitigates this interference, recovering much of the emotion capability while further improving duration accuracy, demonstrating the value of joint refinement.

Building on Joint-RL, IED+GRPO improves content fidelity, MER-based emotion accuracy, and speaking-rate direction accuracy, but reduces exact-duration precision, with no corresponding emotion gain under the independent E2V evaluator. This suggests a shift in the balance among control objectives rather than uniform improvement. Across stages, speaker similarity remains relatively stable, while boundary pitch discontinuity varies modestly.

\begin{table}[t]
\caption{Objective evaluation of the emotion counterfactual test.}
\label{tab:emotion_counterfactual}
\renewcommand{\arraystretch}{1.1}
\centering
\resizebox{0.48\textwidth}{!}{%
\begin{tabular}{l|cccccccc}
\hline
\multirow{2}{*}{System} & \multicolumn{2}{c}{MER$\uparrow$} & \multicolumn{2}{c}{E2V$\uparrow$} & \multicolumn{2}{c}{Dual$\uparrow$} & \multirow{2}{*}{ASR$\downarrow$} & \multirow{2}{*}{Spk$\uparrow$} \\
                        & Acc              & Utt            & Acc         & Utt                 & Acc              & Utt             &                                  &                                \\ \hline
CosyVoice2              & 25.6        & 4.8            & 23.0          & 5.2                 & 16.8             & 1.2             & 30.61                            & 0.748                          \\
CosyVoice3              & 35.4         & 6.4            & 24.0          & 3.6                 & 14.6
& 0.0               & 1.43                             & \textbf{0.780}                  \\
IndexTTS2               & 28.2         & 1.2            & 24.8        & 2.4     & 18.0               & 0.0               & \textbf{1.35}                    & 0.706                          \\
TED-TTS                 & 48.8        & 13.6           & \textbf{41.4}        & 13.2                & 32.0               & 6.8             & 1.37                             & 0.540                           \\ \hline
Joint-RL                & \textbf{71.4}    & \textbf{52.0}    & 40.8        & \textbf{16.8}       & \textbf{36.4}    & \textbf{12.4}   & 1.40                              & 0.767                          \\ \hline
\end{tabular}%
}
\vspace{-5mm}
\end{table}

\begin{table}[t]
\caption{Objective evaluation of the duration counterfactual test.}
\label{tab:duration_counterfactual}
\centering
\begingroup
\fontsize{9pt}{10.8pt}\selectfont
\setlength{\tabcolsep}{1pt}
\renewcommand{\arraystretch}{0.96}

\begin{tabularx}{\columnwidth}
{@{}L{0.2\columnwidth}*{6}{Y}@{}}
\toprule
System
& ASR$\downarrow$
& Spk$\uparrow$
& A15$\uparrow$
& U15$\uparrow$
& RA$\uparrow$
& DA$\uparrow$ \\
\midrule
CosyVoice2 & 74.6 & 0.865 & 25.3 & 4.7 & 13.0 & 34.7 \\
CosyVoice3 & 3.5 & \textbf{0.875} & 43.3 & 18.0 & 26.3 & 41.7 \\
TED-TTS & 6.0 & 0.776 & 47.0 & 18.0 & 25.7 & 40.7 \\
MAGIC-TTS & 41.9 & 0.764 & \textbf{51.3} & \textbf{28.0} & 17.3 & 43.3 \\
\midrule
Joint-RL & \textbf{3.0} & 0.863 & 50.3 & 26.7 & \textbf{27.7} & \textbf{45.7} \\
\bottomrule
\end{tabularx}
\endgroup
\vspace{-5mm}
\end{table}

\subsection{Counterfactual Evaluation}
Tables~\ref{tab:emotion_counterfactual} and~\ref{tab:duration_counterfactual} present the counterfactual results. On neutral-content emotion requests, Joint-RL achieves the highest utterance-level accuracy under both classifiers, although TED-TTS slightly outperforms it in E2V segment accuracy. For duration control, Joint-RL achieves accuracy comparable to the explicitly conditioned MAGIC-TTS while maintaining much lower ASR error. It also leads in range and direction accuracy, demonstrating more reliable adherence to qualitative speaking-rate instructions.

\subsection{Instruction Robustness}

\begin{table}[t]
\caption{Evaluation under unseen instructions.}
\label{tab:unseen_instruct_formats}
\renewcommand{\arraystretch}{1.1}
\centering
\resizebox{0.48\textwidth}{!}{%
\begin{tabular}{c|ccccccc}
\hline
\multirow{2}{*}{System} & \multirow{2}{*}{ASR$\downarrow$} & \multicolumn{1}{c}{MER$\uparrow$} & \multicolumn{1}{c}{E2V$\uparrow$} & \multirow{2}{*}{A15$\uparrow$/U15$\uparrow$} & \multirow{2}{*}{RA$\uparrow$/DA$\uparrow$} & \multicolumn{2}{c}{Joint}     \\
                        &                                  & ACC/UTT                           & ACC/UTT                           &                                              &                                            & MER$\uparrow$ & E2V$\uparrow$ \\ \hline
CosyVoice3              & 35.3                             & 27.6/3.5                          & 25.8/3.5                          & 28.4/7.2                                     & 21.1/40.2                                  & 0.0           & 0.0           \\
TED-TTS                 & 7.1                              & \textbf{51.8/19.4}                & \textbf{47.1/14.7}                & \textbf{46.9}/16.5                           & \textbf{26.3/49.0}                         & 1.2  & 0.6  \\
MAGIC-TTS               & 29.2                             & -                                 & -                                 & 46.4/\textbf{21.7}                                    & 22.2/47.0                         & -             & -             \\
Joint-RL                & 19.8                             & 35.3/8.2                          & 31.1/2.9                          & 32.0/12.4                                    & 19.6/32.5                                  & 1.2           & 0.6           \\
IED+GRPO                & \textbf{3.6}                     & 44/12.4                           & 40.4/7.1                          & 41.8/10.3                                    & 20.6/33.5                                  & \textbf{2.4}  & \textbf{1.2}  \\ \hline
\end{tabular}%
}
\vspace{-5mm}
\end{table}

Table~\ref{tab:unseen_instruct_formats} evaluates robustness to paraphrased instructions. IED+GRPO improves content fidelity, emotion, segment-level duration, and joint accuracy over Joint-RL, though strict duration accuracy remains limited. Together with standard-control results, this highlights a robustness--precision trade-off. Unlike structured-input baselines, our model remains competitive while handling natural-language variations.


\subsection{Human Evaluation.}

\begin{table}[t]
\caption{Human evaluation of emotion perception and speech naturalness.
A: Agreement with human votes (All), majority labels (Majority),
and counterfactual votes (CF); Requested label denotes the target emotion.
B: Boundary/Overall MOS (1--5) and audible seam rate (\%).}
\label{tab:human_combined}
\renewcommand{\arraystretch}{1.0}
\centering
\resizebox{0.48\textwidth}{!}{%
\begin{tabular}{lccc}
\hline
\multicolumn{4}{l}{\textbf{A. Classifier-disagreement listening test}}                                                         \\ \hline
\multicolumn{1}{l|}{\multirow{2}{*}{Reference}} & All votes$\uparrow$ & Majority$\uparrow$ & CF votes$\uparrow$                \\
\multicolumn{1}{l|}{}                           & $n=360$             & $n=107$            & $n=81$                            \\ \hline
\multicolumn{1}{l|}{MERaLiON}                   & 45.56               & 55.14              & 40.74                             \\
\multicolumn{1}{l|}{emotion2vec+}               & 33.06               & 28.97              & 25.93                             \\
\multicolumn{1}{l|}{Requested label}            & 57.22               & 70.09              & 38.27                             \\ \hline
\multicolumn{4}{l}{\textbf{B. Naturalness listening test}}                                                                     \\ \hline
\multicolumn{1}{l|}{\multirow{2}{*}{System}}    & Boundary            & Overall            & \multirow{2}{*}{Seam$\downarrow$} \\
\multicolumn{1}{l|}{}                           & MOS$\uparrow$       & MOS$\uparrow$      &                                   \\ \hline
\multicolumn{1}{l|}{CosyVoice3 (concat.)}       & 3.34                & 3.38               & 41.67                             \\
\multicolumn{1}{l|}{IndexTTS2 (concat.)}        & 3.99                & 3.98               & 9.17                              \\
\multicolumn{1}{l|}{Joint-RL (one pass)}        & 4.64                & 4.63               & 2.50                              \\ \hline
\end{tabular}%
}
\vspace{-5mm}
\end{table}

Table~\ref{tab:human_combined} presents the human evaluation results. On classifier-disagreement samples, MERaLiON aligns more closely with human judgments than emotion2vec+, suggesting that the observed gains are not solely attributable to reward-classifier overfitting, without implying overall classifier superiority. Moreover, majority human judgments generally favor the requested emotion, supporting perceptual control effectiveness. In naturalness evaluation, Joint-RL achieves higher boundary and overall MOS with substantially fewer audible seams than concatenation-based baselines, demonstrating smoother transitions and more coherent speech.

\section{Conclusion}
In this work, we proposed a unified post-training framework for natural-language emotion and duration control at the segment level, without inference-time modules. Staged SFT and GRPO outperform most baselines, with counterfactual, robustness, and human evaluations validating control effectiveness and naturalness.

\section{Compliance with Ethical Standards}
The authors' institution does not provide a formal ethics review procedure
applicable to non-biomedical, minimal-risk listening studies of this nature.
The study was conducted in accordance with the principles of the Declaration
of Helsinki. Participants were informed of the research purpose and
voluntarily agreed to participate. No personally identifiable or sensitive
information was collected.

\section{Acknowledgments}
No funding was received for conducting this study. The authors have
no relevant financial or nonfinancial interests to disclose.
\bibliographystyle{IEEEbib}
\bibliography{strings,refs}

@article{du2024cosyvoice,
  title={Cosyvoice 2: Scalable streaming speech synthesis with large language models},
  author={Du, Zhihao and Wang, Yuxuan and Chen, Qian and Shi, Xian and Lv, Xiang and Zhao, Tianyu and Gao, Zhifu and Yang, Yexin and Gao, Changfeng and Wang, Hui and others},
  journal={arXiv preprint arXiv:2412.10117},
  year={2024}
}

@inproceedings{zhou2026indextts2,
  title={Indextts2: A breakthrough in emotionally expressive and duration-controlled auto-regressive zero-shot text-to-speech},
  author={Zhou, Siyi and Zhou, Yiquan and He, Yi and Zhou, Xun and Wang, Jinchao and Deng, Wei and Shu, Jingchen},
  booktitle={Proceedings of the AAAI Conference on Artificial Intelligence},
  volume={40},
  pages={35139--35148},
  year={2026}
}

@article{chen2026flexivoice,
  title={Flexivoice: Enabling flexible style control in zero-shot tts with natural language instructions},
  author={Chen, Dekun and Zhang, Xueyao and Wang, Yuancheng and Dai, Kenan and Ma, Li and Wu, Zhizheng},
  journal={arXiv preprint arXiv:2601.04656},
  year={2026}
}

@inproceedings{guo2023prompttts,
  title={Prompttts: Controllable text-to-speech with text descriptions},
  author={Guo, Zhifang and Leng, Yichong and Wu, Yihan and Zhao, Sheng and Tan, Xu},
  booktitle={ICASSP 2023 IEEE International Conference on Acoustics, Speech and Signal Processing (ICASSP)},
  pages={1--5},
  year={2023},
  organization={IEEE}
}

@article{yang2024instructtts,
  title={Instructtts: Modelling expressive tts in discrete latent space with natural language style prompt},
  author={Yang, Dongchao and Liu, Songxiang and Huang, Rongjie and Weng, Chao and Meng, Helen},
  journal={IEEE/ACM Transactions on Audio, Speech, and Language Processing},
  volume={32},
  pages={2913--2925},
  year={2024},
  publisher={IEEE}
}

@article{wang2026word,
  title={Word-level emotional expression control in zero-shot text-to-speech synthesis},
  author={Wang, Tianrui and Wang, Haoyu and Ge, Meng and Gong, Cheng and Qiang, Chunyu and Ma, Ziyang and Huang, Zikang and Yang, Guanrou and Wang, Xiaobao and Chng, Eng-Siong and others},
  journal={Advances in Neural Information Processing Systems},
  volume={38},
  pages={147377--147405},
  year={2026}
}

@inproceedings{liang2026ted,
  title={TED-TTS: Training-Free Intra-Utterance Emotion and Duration Control for Text-to-Speech Synthesis},
  author={Liang, Qifan and Liu, Yuansen and Wei, Ruixin and Lu, Nan and Zhao, Junchuan and Wang, Ye},
  booktitle={Proceedings of the 64th Annual Meeting of the Association for Computational Linguistics (Volume 1: Long Papers)},
  pages={23485--23508},
  year={2026}
}

@article{mai2026magic,
  title={Magic-tts: Fine-grained controllable speech synthesis with explicit local duration and pause control},
  author={Mai, Jialong and Xing, Xiaofen and Xu, Xiangmin},
  journal={arXiv preprint arXiv:2604.21164},
  year={2026}
}

@inproceedings{liu2026group,
  title={Group relative policy optimization for text-to-speech with large language models},
  author={Liu, Chang and Hu, Ya-Jun and Gao, Ying-Ying and Zhang, Shi-Lei and Ling, Zhen-Hua},
  booktitle={ICASSP 2026 IEEE International Conference on Acoustics, Speech and Signal Processing (ICASSP)},
  pages={16132--16136},
  year={2026},
  organization={IEEE}
}

@inproceedings{li2026emorl,
  title={Emorl-tts: Reinforcement learning for fine-grained emotion control in llm-based tts},
  author={Li, Haoxun and Liu, Yu and Sun, Yuqing and Shi, Hanlei and Qu, Leyuan and Li, Taihao},
  booktitle={ICASSP 2026 IEEE International Conference on Acoustics, Speech and Signal Processing (ICASSP)},
  pages={17272--17276},
  year={2026},
  organization={IEEE}
}

@article{liu2026emotra,
  title={EmoTra-TTS: Smooth Intra-Utterance Emotion Transitions for Speech Synthesis},
  author={Liu, Tianchi and Song, Zeyang and Wang, Tianrui and Li, Zhipeng and Xu, Chenglin and Guo, Yiwen},
  journal={arXiv preprint arXiv:2608.23791},
  year={2026}
}

@inproceedings{kim2016sequence,
  title={Sequence-level knowledge distillation},
  author={Kim, Yoon and Rush, Alexander M},
  booktitle={Proceedings of the 2016 conference on empirical methods in natural language processing},
  pages={1317--1327},
  year={2016}
}

@article{asaria2026reliable,
  title={Reliable Neural-Codec Text-to-Speech by ASR Self-Verification and Distillation: Near-Zero Catastrophic Failures Across Models and Codecs},
  author={Asaria, Ali and Salomone, Tony and Gandhi, Deep},
  journal={arXiv preprint arXiv:2606.18323},
  year={2026}
}

@article{zhou2026sequential,
  title={Sequential Trajectories and Simultaneous Blending: Multi-Emotion Modeling for Instruction-Following TTS},
  author={Zhou, Yan and Hong, Yun and Feng, Yang},
  journal={arXiv preprint arXiv:2608.30325},
  year={2026}
}

@inproceedings{zhou2021seen,
  title={Seen and unseen emotional style transfer for voice conversion with a new emotional speech dataset},
  author={Zhou, Kun and Sisman, Berrak and Liu, Rui and Li, Haizhou},
  booktitle={ICASSP 2021 IEEE International Conference on Acoustics, Speech and Signal Processing (ICASSP)},
  pages={920--924},
  year={2021},
  organization={IEEE}
}

@article{anastassiou2406seed,
  title={Seed-tts: A family of high-quality versatile speech generation models},
  author={Anastassiou, Philip and Chen, Jiawei and Chen, Jitong and Chen, Yuanzhe and Chen, Zhuo and Chen, Ziyi and Cong, Jian and Deng, Lelai and Ding, Chuang and Gao, Lu and others},
  journal={URL https://arxiv.org/abs/2406.02430},
  year={2024}
}

@article{sailor2025meralion,
  title={MERaLiON-SER: Robust Speech Emotion Recognition Model for English and SEA Languages},
  author={Sailor, Hardik B and Ti, Aw Ai and Nancy, Chen Fang Yih and Lay, Chiu Ying and Yang, Ding and Yingxu, He and Ridong, Jiang and Jingtao, Li and Jingyi, Liao and Zhuohan, Liu and others},
  journal={arXiv preprint arXiv:2511.04914},
  year={2025}
}

@article{shi2026qwen3,
  title={Qwen3-asr technical report},
  author={Shi, Xian and Wang, Xiong and Guo, Zhifang and Wang, Yongqi and Zhang, Pei and Zhang, Xinyu and Guo, Zishan and Hao, Hongkun and Xi, Yu and Yang, Baosong and others},
  journal={arXiv preprint arXiv:2601.21337},
  year={2026}
}

@article{shao2024deepseekmath,
  title={Deepseekmath: Pushing the limits of mathematical reasoning in open language models},
  author={Shao, Zhihong and Wang, Peiyi and Zhu, Qihao and Xu, Runxin and Song, Junxiao and Bi, Xiao and Zhang, Haowei and Zhang, Mingchuan and Li, YK and Wu, Yang and others},
  journal={arXiv preprint arXiv:2402.03300},
  year={2024}
}

@article{du2025cosyvoice,
  title={Cosyvoice 3: Towards in-the-wild speech generation via scaling-up and post-training},
  author={Du, Zhihao and Gao, Changfeng and Wang, Yuxuan and Yu, Fan and Zhao, Tianyu and Wang, Hao and Lv, Xiang and Wang, Hui and Ni, Chongjia and Shi, Xian and others},
  journal={arXiv preprint arXiv:2505.17589},
  year={2025}
}

@inproceedings{ma2024emotion2vec,
  title={emotion2vec: Self-supervised pre-training for speech emotion representation},
  author={Ma, Ziyang and Zheng, Zhisheng and Ye, Jiaxin and Li, Jinchao and Gao, Zhifu and Zhang, Shiliang and Chen, Xie},
  booktitle={Findings of the Association for Computational Linguistics: ACL 2024},
  pages={15747--15760},
  year={2024}
}

@article{an2024funaudiollm,
  title={Funaudiollm: Voice understanding and generation foundation models for natural interaction between humans and llms},
  author={An, Keyu and Chen, Qian and Deng, Chong and Du, Zhihao and Gao, Changfeng and Gao, Zhifu and Gu, Yue and He, Ting and Hu, Hangrui and Hu, Kai and others},
  journal={arXiv preprint arXiv:2407.04051},
  year={2024}
}

@article{koizumi2023libritts,
  title={Libritts-r: A restored multi-speaker text-to-speech corpus},
  author={Koizumi, Yuma and Zen, Heiga and Karita, Shigeki and Ding, Yifan and Yatabe, Kohei and Morioka, Nobuyuki and Bacchiani, Michiel and Zhang, Yu and Han, Wei and Bapna, Ankur},
  journal={arXiv preprint arXiv:2305.18802},
  year={2023}
}

@article{wang2023cam++,
  title={Cam++: A fast and efficient network for speaker verification using context-aware masking},
  author={Wang, Hui and Zheng, Siqi and Chen, Yafeng and Cheng, Luyao and Chen, Qian},
  journal={arXiv preprint arXiv:2303.00332},
  year={2023}
}

\end{document}